# Evaluation of Incremental Forming of Sheet Metal with a Negative Die: Experimental and Numerical Approaches

Alireza Hajfathalian[1], Majid Elyasi[1], Mohammad Javad Mirnia[1]

1- Faculty of Mechanical Engineering, Babol Noshirvani University of Technology, Babol, Mazandaran, Iran

**Abstract**

Two-point incremental forming (TPIF) is a flexible sheet metal forming process, commonly divided into positive and negative configurations. In order to improve the dimensional accuracy, thickness distribution, and overall formability of conical sheet parts, this study examines the negative TPIF approach. A series of experiments were conducted by varying key process parameters including tool diameter, vertical step size, forming strategy, and lubricant type. The forming process was performed on 0.9 mm thick St12 steel sheets using a CNC milling machine equipped with a full die set and a spherical-head forming tool. According to the results, increasing the tool diameter reduced material thinning and failure risk by improving thickness distribution and increasing the minimum wall thickness from 0.24 mm to 0.5 mm. On the other hand, sharp features were better defined with smaller tool diameters. The corresponding plastic strain was decreased and the risk of tearing was reduced by using a multi-stage forming technique, which also improved strain distribution and decreased thinning. Lubricant type showed limited influence on geometric accuracy or thickness uniformity but had a noticeable effect on surface finish. Overall, this study demonstrates that tool geometry and forming strategy are critical in optimizing the performance of negative TPIF processes. The findings provide practical guidelines for improving the structural integrity and precision of incrementally formed sheet metal parts.

# 1. Introduction

Currently, manufacturing industries are seeking flexible methods for producing medium-sized products with better quality. This is due to the significant advancement in technology and market trends favoring lower-volume, higher-quality productions, leading to an increased demand in this regard. Rapid prototyping emerges as the initial and most crucial stage in addressing this need. In the realm of rapid prototyping for sheet metal products, the incremental forming process stands out as a suitable option. This method offers advantages such as reducing the time and cost of the design-to-production cycle, increasing the speed of product design processes, facilitating rapid modifications, and accelerating product introduction to the market [1]. The incremental forming process for sheet metal is broadly divided into two categories: negative or single-point incremental forming and positive or two-point incremental forming[1]. Single-point incremental forming is executed simply and easily, without the need for complex tolerances. This process is carried out without the use of special dies, allowing for quicker adjustments and greater flexibility compared to two-point incremental forming [2]. However, two-point incremental forming offers high flexibility and requires minimal forming force while also providing significant shaping capabilities. This flexibility and shaping capability make it a suitable option for rapid prototyping of sheet metal [3].

In the two-point incremental forming process for sheet metal, a spherical-headed tool is used to locally shape the metal until it reaches the final product shape. One of the most significant advantages of this process is the absence of complex and expensive dies and tools; instead, it can be executed with a simple design and a CNC milling machine. To reduce the required forming force and increase the formability range of the sheet metal, a small-sized tool is employed in this process. This action concentrates the deformation operation on a small area of the sheet, thereby reducing potential damage to the final product. Due to the gradual and localized forming of the sheet, the forming force is concentrated in the tool-contact area, leading to increased formability and flexibility of the process. However, thinning is one of the drawbacks of this process, which can result in reduced strength of the final product [4]. The two-point incremental forming process has less spring back compared to the single-point incremental forming process, which results in increased geometric accuracy [5].

The dies used in the two-point incremental forming process are classified into two main categories: partial and complete dies. Partial dies offer greater flexibility in shaping parts with different geometries compared to complete dies [6]. In complete dies, the sheet remains in contact with the die throughout the forming process, which results in significantly higher precision compared to partial dies. However, some drawbacks of the two-point incremental forming process with complete dies include limited flexibility and time-consuming nature due to the precise part production. Additionally, two-point incremental forming with complete dies is much more costly than with partial dies [7]. In the field of incremental forming, numerous research studies have been conducted, some of which are briefly mentioned below. Nikdooz et al [4] investigated the formability of aluminum truncated cones with a 70-degree angle using the incremental forming process in both single and two stages. The results demonstrated that the two-stage strategy led to a twofold improvement in the minimum thickness compared to the single-stage approach.

Ham and Jeswiet [8] conducted experimental tests on aluminum sheets(AA3003). They performed their experiments at two speeds: 600 and 1000 revolutions per minute (RPM). They concluded that at 600 RPM, the formability is higher, attributing this increase in formability to the generation of more frictional heat in the contact area between the tool and the sheet. They also found that reducing the feed rate in incremental forming improves formability. Taherkhani et al. [9] conducted investigations in the incremental forming process. They examined the effects of parameters such as initial sheet thickness, tool diameter, and feed rate on the dimensional accuracy and surface roughness of aluminum sheets. Ultimately, they found that reducing the initial sheet thickness and increasing the tool diameter resulted in a reduction in surface roughness of the workpiece, leading to a smoother surface finish in the final part.

Vahdati et al. [10] investigated spring back and its effects on geometry and dimensional accuracy through the incremental forming process, both experimentally and numerically. They developed an analytical model by selecting an appropriate method for process parameters and reducing spring back. They found that increasing the tool diameter, feed rate, spindle speed, and sheet thickness, while decreasing the vertical stage size of the tool, resulted in a reduction in spring back. Manko and Ambrogio [11] examined the effects of tool diameter, wall angle, sheet thickness, and vertical stage size on the changes in minimum thickness induced in the incremental forming process of an aluminum cone. Through the analysis of experimental test results, they concluded that changes in tool diameter have minimal effects on the minimum thickness, but increasing the vertical stage size during the incremental forming process results in an increase in the minimum thickness. Shim and Park [12] conducted research aimed at demonstrating how the forming limit curve behaves in the incremental forming process of sheet metal. By determining and examining the forming limit curve in aluminum, they concluded that the forming limit in the incremental forming process is higher compared to conventional and traditional forming processes. Therefore, the forming limit curve of conventional processes cannot be directly compared and generalized with the incremental forming process. MovahediNia et al. [13] investigated the effect of tube flaring in the incremental forming process. They ultimately concluded that reducing friction between the tool and the sheet and decreasing the feed rate result in increased formability in the incremental forming process.

Kurra and Regalla [14] experimentally and numerically investigated the formability of high-strength steel sheets through the incremental forming process. In this study, cones with variable wall angles, formed by conical, pyramidal, oval, and power generators, were utilized. Ultimately, the results demonstrated that the average maximum wall angle obtained in this study was 75.27 degrees, with a maximum deviation of 4.6 degrees in the wall angle when using various generators. Darzi et al. [15] conducted an experimental investigation on high-temperature incremental forming of AA6061 aluminum alloy. In this study, the applied heating ranged from 25 to 400 degrees Celsius with a uniform distribution throughout the process. They found that the heating employed in this study resulted in a 528% increase in formability during the process. Furthermore, they observed that in terms of the influential factors on formability, temperature ranked as the highest priority, followed by lubrication and vertical stage size.

Ajay et al. [16] conducted research on incremental sheet forming and its various methods. They found that incremental sheet forming can be broadly categorized into two types: single-point incremental forming and two-point incremental forming. Single-point incremental forming offers greater flexibility in the process and has significant potential for reducing tooling costs in prototyping and low-volume production compared to two-point incremental forming. On the other hand, two-point incremental forming exhibits better profile accuracy compared to single-point incremental forming.

Trzepieciński et al. [17] investigated the latest advancements and challenges in the process of forming aluminum sheets and aluminum alloys using single-point incremental forming (SPIF) and two-point incremental forming (TPIF) methods. They concluded that in TPIF, there are two contact points with the sheet surface: one forming tool at the top and a backing tool at the bottom of the sheet. This method, compared to SPIF, offers better formability and geometric accuracy. Employing a multi-stage strategy in TPIF can improve the thickness distribution of the part. Additionally, they experimented with four different tool path strategies in this research. Fabian et al. [18] investigated the influence of two-point incremental forming (TPIF) on the mechanical properties of the produced part, both experimentally and numerically, as residual stresses resulting from the TPIF process significantly affect the mechanical properties of the produced part. They used a polymer pad as a flexible tooling during the TPIF process. The numerical and experimental results of their research indicate the effect of stress superposition on the produced part. The effect of stress superposition in this process was analyzed in terms of process forces, sheet thickness, and material hardness. As a result, they found that the process forces in the x, y, and z directions increase due to elastic deformation with the flexible tooling, and TPIF with flexible tooling leads to a reduction in sheet thickness.

Scheffler et al. [19] investigated the extent to which two-point incremental forming (TPIF) could be used for producing the outer skin and some automotive components. They pursued their objective using numerical and experimental methods and ultimately concluded that employing TPIF for producing the entire body of a car is not practical. This conclusion was drawn because there are simpler and more cost-effective forming processes for curved surfaces, such as rolling. Additionally, TPIF cannot produce the required 90-degree multi-sided flanges at the front of the car. Esmailian et al. [20] investigated the effects of parameters such as vertical stage size, tool diameter, tool rotation speed, and polymer type on the average tool force and thickness uniformity in TPIF. They found that the vertical stage size has the greatest impact on the average tool force, followed by parameters such as tool diameter, polymer type, and tool rotation speed. They also observed that the vertical stage size, polymer type, tool rotation speed, and tool diameter, respectively, have the most significant influence on the change in maximum thickness. Ou et al. [21] investigated the influence of TPIF process parameters on the thickness uniformity of an irregular staircase-shaped part made of AL1060 alloy. They found that the thickness uniformity improves with increasing tool diameter but decreases significantly with increasing feed rate and stage size. Additionally, reducing the forming angle between each pass for the metal flow and improving the accuracy of the formed part are beneficial. Changing the direction of tool movement between adjacent passes significantly improves the thickness uniformity of the wall.

As mentioned, the two-point incremental forming process has gained a special place in the manufacturing industry due to its high flexibility, simplicity, and cost-effectiveness compared to other methods. However, this process still faces challenges. In this study, an innovative method for two-point incremental forming of 0.9 mm thick steel sheets using a negative die has been proposed. This research focuses on improving dimensional accuracy and thickness distribution of the sheet metal, which is of paramount importance. Dimensional accuracy can directly impact the conformity of parts to design requirements, while thickness distribution can affect the strength and durability of the parts. The parameters used in this study include tool diameter (for determining part dimensions and final surface quality), vertical stage (for adjusting forming depth and eliminating irregularities), lubricant type (for reducing friction and improving material flow), and forming strategy (for optimizing the process and increasing efficiency). Proper examination and adjustment of these parameters can ensure the improvement of the final process quality and part performance. Additionally, advanced simulations have been used in this research for modeling and predicting the properties of the produced parts.

It is worth noting that most of the research conducted on two-point incremental forming has focused on rotary or low-flange angle parts. However, this study explores the experimental investigation of two-point incremental forming in a solid negative die with a 47-degree wall angle.

Another noteworthy point is that previous studies rarely examined two-point incremental forming with negative solid dies using single or two-stage forming strategies. However, this research delves into the experimental investigation of two-point incremental forming with a negative solid die using St12 steel sheet along with lubrication in both single and two-stage forming strategies. Ultimately, this research takes an innovative approach to improving the incremental forming process, with experimental results and advanced simulations confirming the significance of this method and providing precise and practical evaluation.

## 2. Experimental procedure

### 2.1. Mechanical Properties

As mentioned earlier, ST12 steel is used in this study. In the next stage, the mechanical properties of the ST12 steel sheet must be introduced into the Abaqus software. These properties are crucial for simulating the progressive forming process in Abaqus and are divided into two parts: elastic and plastic. The elastic mechanical properties include the elastic modulus, density, Poisson's ratio, and yield stress in the elastic region. These properties will be entered into the software according to Table 1.

Table 1: Mechanical Properties of ST12 Steel Sheet

| Density (g/cm³) | 7.87 |
|---|---|
| Elastic Modulus (Gpa)[22] | 202 |
| Yield Stress (MPa) | 195 |
| Poisson's Ratio | 0/3 |

It is worth mentioning that for the plastic region of the metal sheet up to the necking point in this study, the Swift relationship (Equation 1), known as the Swift work hardening equation, is used to describe the material's behavior before the necking phenomenon in the specimen occurs. Due to the necking phenomenon observed in the uniaxial tensile test, the plastic deformation behavior until the sheet's fracture, which occurs after the necking point, is not uniform. Therefore, the stress-strain data up to the necking point is extrapolated using work hardening.

$$\bar{\sigma} = K(\varepsilon_0 + \bar{\varepsilon}^p)^n \qquad \bar{\varepsilon}^p \leq \bar{\varepsilon}^p_{necking} \qquad (1)$$

the parameters K, $\varepsilon_0$, $\bar{\varepsilon}^p$, $\bar{\varepsilon}^p_{necking}$, and n represent the hardening parameter, pre-strain, strain hardening exponent, necking strain, and strain hardening exponent, respectively. These parameters are crucial for characterizing the plastic behavior of ST12 steel sheet using the Swift hardening equation.

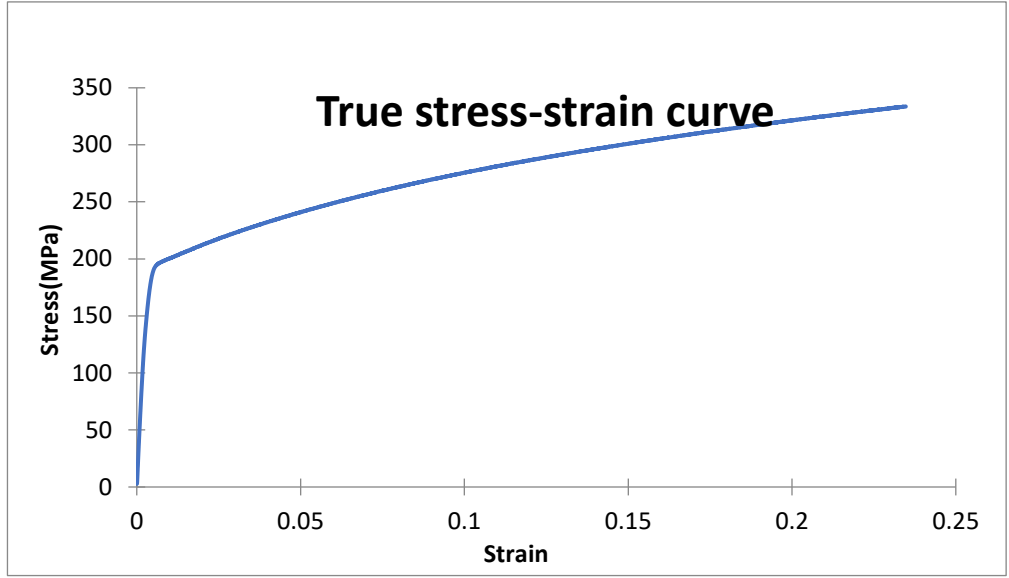

Figure 1 shows the true stress-strain curve up to the necking point for the St12 sheet.

It is worth mentioning that for the plastic region of the metal sheet up to the necking point, the Swift equation (Eq. 1) has been used in this study, is known as the Swift hardening law. This equation describes the material behavior before the onset of necking in the sample. Due to the necking of the sample in the uniaxial tensile test, the plastic deformation behavior up to the point of rupture, which occurs after necking, is not uniform. Therefore, stress-strain data up to the necking point are extrapolated using the hardening law[23].

$$r = \frac{\varepsilon_{22}}{\varepsilon_{33}} \qquad (2)$$

Table 2: Values of the Parameters for the Swift Hardening Equation and the Elastic Region

| Parameters | Unit | Values |
| --- | --- | --- |
| Yield strength coefficient (K) | MPa | 479.2 |
| Strain hardening exponent (n) | --- | 0.271 |
| $\varepsilon_0$ | --- | 0.036 |

| $\bar{\varepsilon}^p_{necking}$ | --- | 0.022 |

$\varepsilon_{22}$ and $\varepsilon_{33}$ in Eq. 2 represent the lateral and thickness strains, respectively. The anisotropy coefficient is calculated by substituting the relationships for lateral and thickness strain into Eq. 3. It should be noted that in materials where the mechanical properties are isotropic, meaning uniform in all directions, the values of thickness and lateral strains are equal, and the anisotropy coefficient is equal to one.

$$r = \frac{\ln\frac{w}{w_0}}{\ln\frac{t}{t_0}} \qquad (3)$$

w and $w_0$ respectively denote the width of the sample after tension in uniaxial tensile testing and the initial width of the sample before tension. Similarly, t and $t_0$ represent the thickness of the sheet after tension in uniaxial tensile testing and the initial thickness of the sheet. Due to measurement errors in calculating thickness strain, Eq. 3 is rewritten in terms of longitudinal and lateral strains (assuming volume constancy) as equations (4) to (6) [23].

$$\varepsilon_{11} + \varepsilon_{22} + \varepsilon_{33} = 0 \qquad (4)$$

$$r = -\frac{\varepsilon_{22}}{\varepsilon_{11} + \varepsilon_{22}} \qquad (5)$$

$$r = \frac{-\ln\frac{w}{w_0}}{\ln\frac{l}{l_0}+\ln\frac{w}{w_0}} \qquad (6)$$

In equations (4 to 6), $\varepsilon_{11}$, l, and $l_0$ respectively represent the longitudinal strain, length of the sample after tension, and initial length of the sample before tension. The angle between the direction of the cut of the tensile sample relative to the rolling direction plays a significant role in determining the anisotropy coefficient. The angle α denotes the angle between the direction of the cut of the tensile sample relative to the rolling direction, and its measurement method is illustrated in Figure 2. According to Figure 2, the anisotropy coefficient is examined at three angles: 0, 45, and 90 degrees relative to the rolling direction[23].

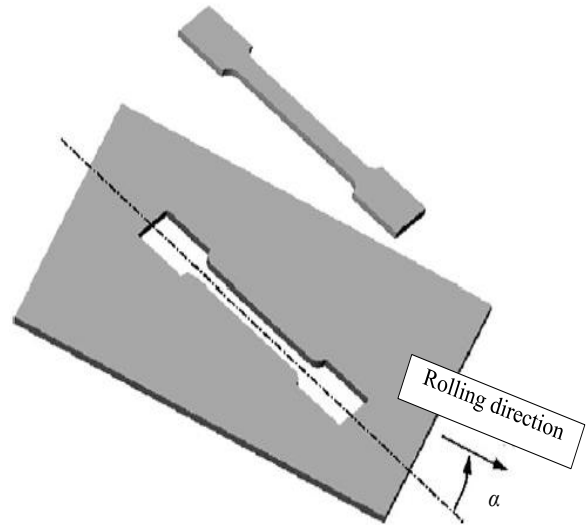

Figure 2 depicts the orientation of the specimen relative to the rolling direction[23].

Lankford coefficients are used as an indicator of the internal anisotropy of a material, demonstrating the difference in anisotropy in various directions of a sample. To calculate these coefficients, uniaxial tensile samples are cut in three different directions (rolling direction, diagonal, and perpendicular to the rolling direction) using a wire cutting machine. After conducting tensile tests, the transverse and longitudinal strains of the samples are calculated. Then, using the average width of the sample at three different points after the tensile test and the initial width, the transverse strain is determined. Subsequently, using equations (2) to (6), the r-values in three different directions (rolling direction, diagonal, and perpendicular to the rolling direction) are calculated. $r_0$, $r_{45}$, $r_{90}$ are the anisotropy coefficients at 0, 45, and 90 degrees, respectively. Based on the discussed materials and the results obtained from the experimental tests, the anisotropy coefficients of the St12 steel sheet are shown in Table 3.

Table 3: Anisotropy Coefficients of the Steel Sheet

| $r_0$ | $r_{45}$ | $r_{90}$ |
|---|---|---|
| 1.527 | 1.03 | 1.235 |

Based on the experimental results, the Hill 48 yield criterion is used to study the plastic behavior of the sheet. Subsequently, using these coefficients, the Lankford coefficients for input into Abaqus software are determined, as shown in Table 4[24].

It is worth mentioning that the values of $R_{11}$، $R_{13}$ و $R_{23}$ are all equal to 1, and these characteristics are applied to the material as homogeneous and solid.

Table 4: Lankford Coefficients for Abaqus

| $R_{22}$ | $R_{12}$ | $R_{33}$ |
|---|---|---|
| 0/95 | 0/91 | 1.06 |

## 2.2 Experimental Equipment

In this study, two-point incremental forming was investigated for producing a truncated cone. It is noteworthy that this research was conducted at room temperature. The geometry and dimensions of the final part are shown in Figure 1. All dimensions shown in Figure 3 are in mm. The sloping wall angle of the part was set to a constant value of 47 degrees. The sheet used in this process was a circular St12 steel sheet with a thickness of 0.9 mm and a diameter of 100 mm.

The required equipment for this research included a simple spherical-headed tool, a rigid complete die, and a three-axis numerically controlled milling machine. Figure 4 shows the spherical-headed tool mounted on the tool holder. The spherical-headed tool used in the process was made of tungsten carbide and was selected with diameters of 10, 15, and 20 mm. The equipment set used in this process, as shown in Figures 5-a and 5-b, consisted of a rigid complete die and a sheet holder to ensure that the sheet had zero degrees of freedom in all axes except the Z-axis during the process. The die set included the sheet, sheet holder, rigid complete die, and four M12 screws.

To perform the two-point incremental forming process, the sheet was first cut into a circular shape with a diameter of 100 mm and then placed in the die. After placing the sheet in the die and fastening the sheet holder onto the die with screws, the forming process was initiated using the numerically controlled milling machine. The final part was designed using CATIA software, and then the G-code of the designed final shape was extracted using PowerMill software. This G-code was used to guide the movement of the spherical-headed tool during forming.

In this study, the tool path was in a spiral pattern with vertical stages of 0.5, 1, and 1.5 mm. The tool feed rate was set at 1200 mm per minute, and the spindle speed was 300 revolutions per minute. In the next stage, the die set was clamped onto the table of the numerically controlled machine using clamps, and then the sheet forming operation was carried out with the numerically controlled machine.

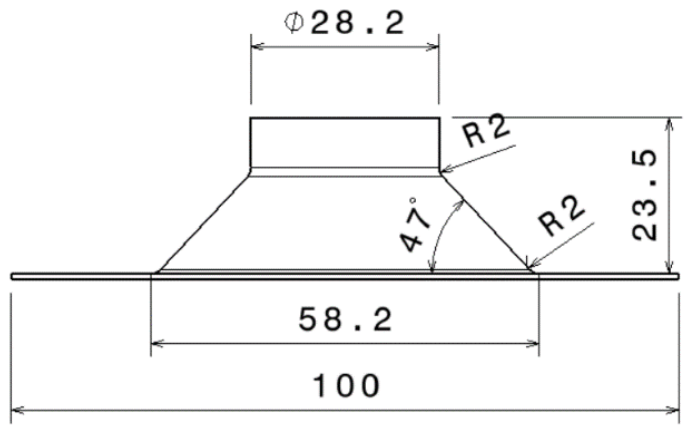

Fig.3 View of the dimensions and shape of the final sample

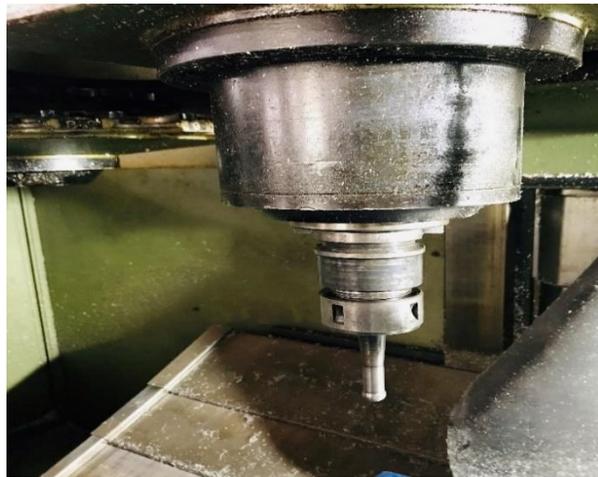

Fig.4 The spherical head tool is placed on the tool holder and spindle of the machine

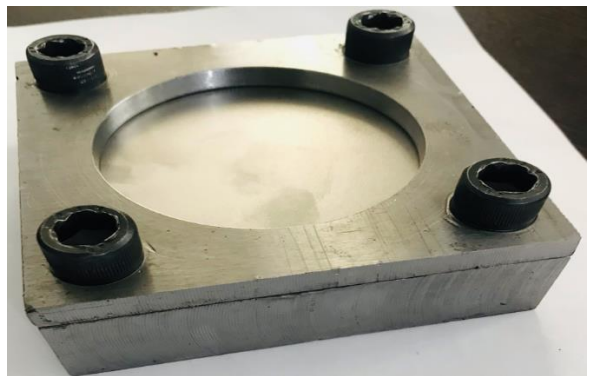

**(a)**

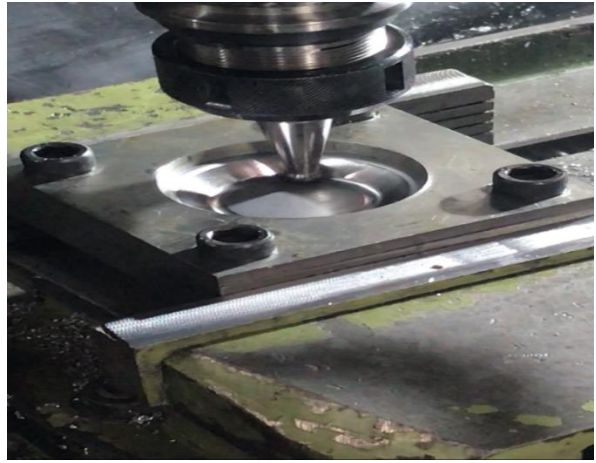

**(b)**
Fig 5 a) completed die b) The equipment used in this research

## 3. Finite element method

The simulation process in this study was carried out using Abaqus software. Given that the issue at hand pertains to metal forming processes, which exhibit nonlinear behavior, the problem in this study falls under the category of quasi-static processes. Consequently, the Abaqus/Explicit solver was utilized for the simulation due to the large deformations present in the TPIF process. Since TPIF is not an axisymmetric process, it is necessary to model the entire sample rather than just a section of it. Therefore, according to the provided explanations, the three main components of the process-tool, sheet, and die-are modeled in Abaqus.

The sheet was meshed as a deformable body using 5452 S4R shell elements, which have 5 integration points. The tool and the die beneath the sheet, due to their rigidity, the reduction in analysis time, and the irrelevance of their mechanical properties, were defined as analytically rigid and three-dimensional shells. For rigid bodies in Abaqus, a reference point can be assigned that represents the entire rigid body. This means that any displacement or boundary condition that needs to be defined for the rigid body is actually applied to the reference point of that rigid body. Therefore, for applying boundary conditions to the tool and die, reference points were defined for them. To simulate the presence of the sheet holder in the process, the degrees of freedom around the edge of the sheet were set to zero throughout the process. The meshing of the simulated model and the method of applying boundary conditions to the tool, die, and sheet are shown in Figure 6.

Another noteworthy point about the process simulation is the meshing method used for the sheet metal used in the research. The mesh size in the meshing of the steel sheet is a very important matter. Its importance is that if the mesh size is taken larger than usual, the results will not be accurate in the end, and if the mesh size is taken smaller than usual and the sheet metal is meshed in detail, it will cause the process time to be very long. Therefore, in this research, mesh convergence was used, which finally converged to a mesh size of 1 mm.

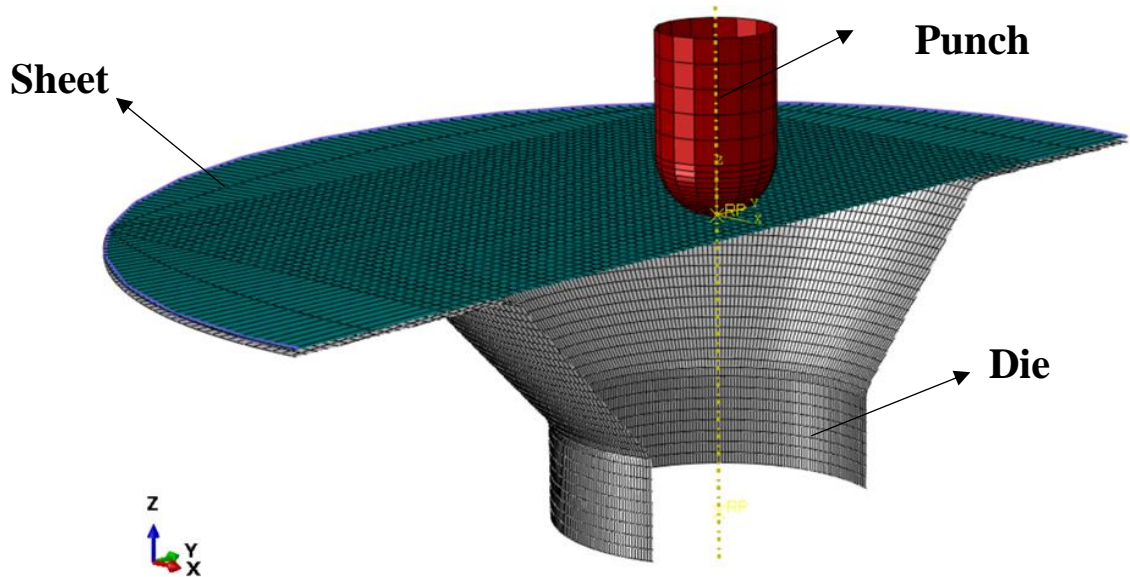

Figure 6: Finite element model of TPIF St12 sheet with negative die

Considering the nature of the process conducted at room temperature and the significant deformation occurring throughout the process, an explicit dynamic solver is chosen for simulating this process. The number of stages defined in the simulation varies for each experimental test and depends on the forming strategy. Therefore, to save time, the analysis time is minimized using MATLAB software, which will be explained further below. In this study, there are interactions between the tool and the sheet, as well as between the sheet and the die. All contacts in the process are defined as surface-to-surface contacts.

Contact conditions are defined mechanically with frictional behavior, and the type of contact between the tool and the sheet, as well as between the sheet and the die, is defined as penalty contact. In this research, the coefficient of friction between the tool and the sheet, as well as between the sheet and the die, is considered as 0.3 based on reference[25]. The sheet and the die must be fully constrained throughout the process, with zero degrees of freedom, to prevent any movement or rotation in any direction. Therefore, the environment of the sheet and the reference point defined for the die are constrained during the process.

However, the reference point for the tool must remain free in all directions throughout the process to ensure proper forming. The path of the tool in Abaqus is defined as follows: first, the die geometry is designed in CATIA software, and then the corresponding design is transferred to PowerMill software to extract G-Code. The extracted G-Code file cannot be directly transferred to Abaqus because Abaqus operates with time-dependent coordinates (X, Y, Z). For this reason, MATLAB software is used to convert the G-Code into time-dependent coordinates. Finally, these time-dependent coordinates are imported into Abaqus and applied to the reference point of the tool. A specific path is defined for comparison of results in this study, as depicted in Figure 7. This path starts from the center of the deformed sample and extends to the outer edge of the sheet.

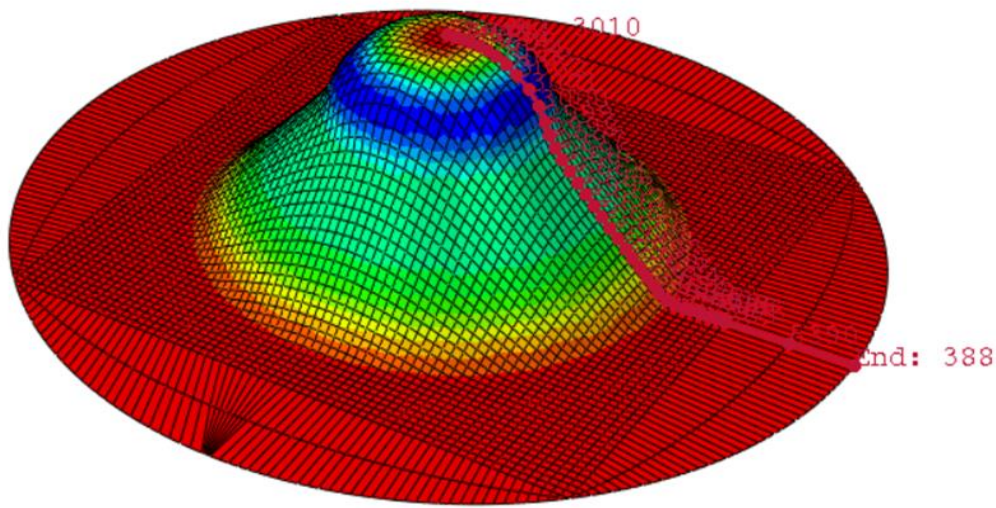

Figure 7: Defined path for comparison of simulated and experimental results

## 4. Results and Discussion

Through multiple experiments and repeatability tests, it was found that the research sheet in this study fractures at a height of 24 mm. Therefore, experiments were conducted according to Figure 3 up to a height of 23.5 mm. Table 5 presents the experiments and their corresponding parameters. In these experiments, a helical motion strategy was employed for forming operations.

Table 5 Experiments and variable parameters in each experiment

| Experiments | Lubricant | Tool diameter (mm) | Vertical stage (mm) | Forming strategy |
|---|---|---|---|---|
| 1 | None | 10 | 0.5 | single-stage |
| 2 | None | 15 | 0.5 | single-stage |
| 3 | None | 15 | 1 | single-stage |
| 4 | None | 15 | 1.5 | single-stage |
| 5 | None | 20 | 0.5 | single-stage |
| 6 | None | 10 | 0.5 | single-stage |
| 7 | None | 10 | 0.5 | two-stage |

## 4.1 Impact of Tool Diameter on Constant Vertical Stage Thickness Distribution

In this stage of the study, the thickness distribution of the sheet is compared in experiments 1, 2, and 5, where the vertical stage is constant (5.0 mm) and the tool diameter is 10, 15, and 20 mm, respectively.

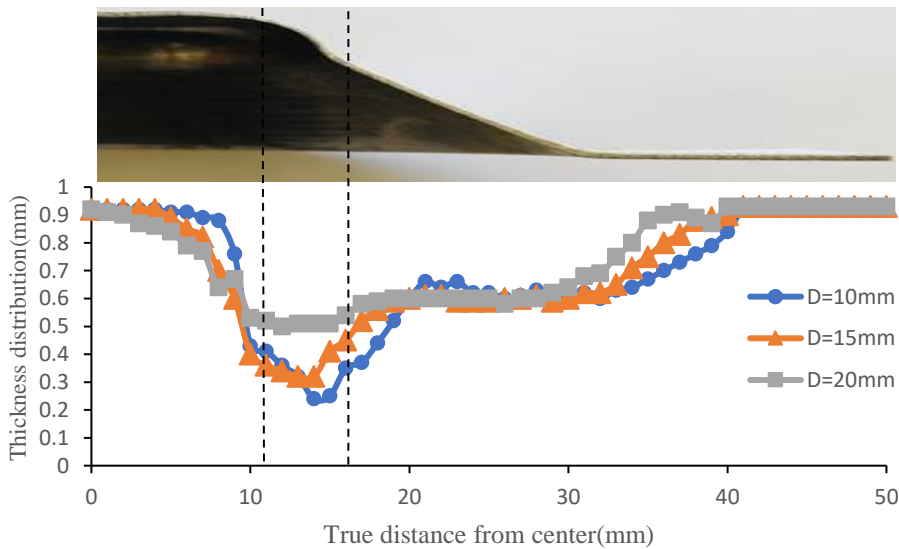

Fig.8 The diagram of the effect of the tool diameter on the thickness distribution

Figure 8 shows the thickness distribution in three experimental samples using tools of different diameters. The minimum thickness for the samples produced with tool diameters of 10 mm, 15 mm, and 20 mm are 0.24 mm, 0.32 mm, and 0.5 mm, respectively. According to the observations and analyses conducted, it has been found that as the tool diameter increases, the contact area between the tool and the sheet for forming increases. This results in the forming forces being distributed over a larger area of the sheet during forming, which reduces the pressure exerted by the tool on the sheet and increases the minimum thickness.

It is worth noting that with smaller tool diameters, the forming becomes more localized, applying greater thickness and longitudinal strains to the sheet. Therefore, in this study, forming with a 20 mm tool diameter exhibits the best thickness distribution. It's important to mention that the thickness distribution was measured using a thickness gauge caliper.

## 4.2 Evaluation of the Tool Diameter's Effect on Vertical Stage Accuracy

In this phase of the research, dimensional accuracy of the produced samples in experiments 1, 2, and 5 is compared, where the vertical stage is fixed at 0.50 mm and the tool diameter varies at 10 mm, 15 mm, and 20 mm, respectively. As depicted in Figures 9, 10, and 11, reducing the tool diameter results in better shaping of the sample corners, consequently enhancing dimensional accuracy of the part. Figure 12 compares the produced samples using tools with different diameters against the designed sample.

Reducing the tool diameter decreases the contact area between the tool and the sheet, thereby improving dimensional accuracy, especially in parts with sharp corners. It is important to note that dimensional accuracy of the samples was measured using a profile projector.

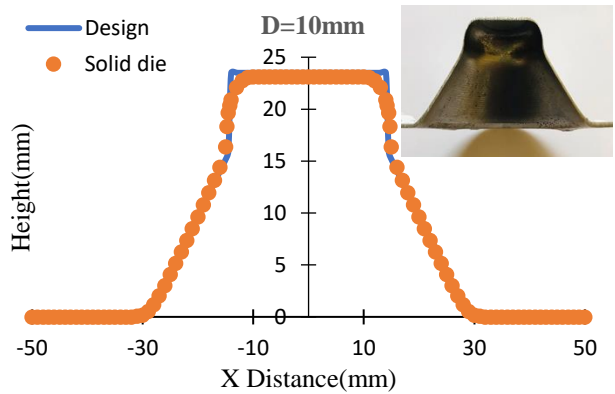

Fig 9 Comparison diagram of the effect of tool diameter of 10 mm on the dimensional accuracy of the production sample with the designed sample

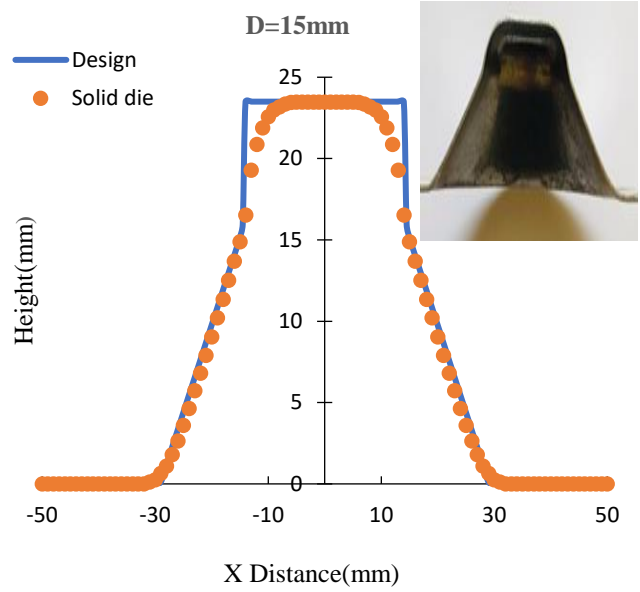

Fig.10 Comparison diagram of the effect of tool diameter of 15 mm on the dimensional accuracy of the production sample with the designed sample

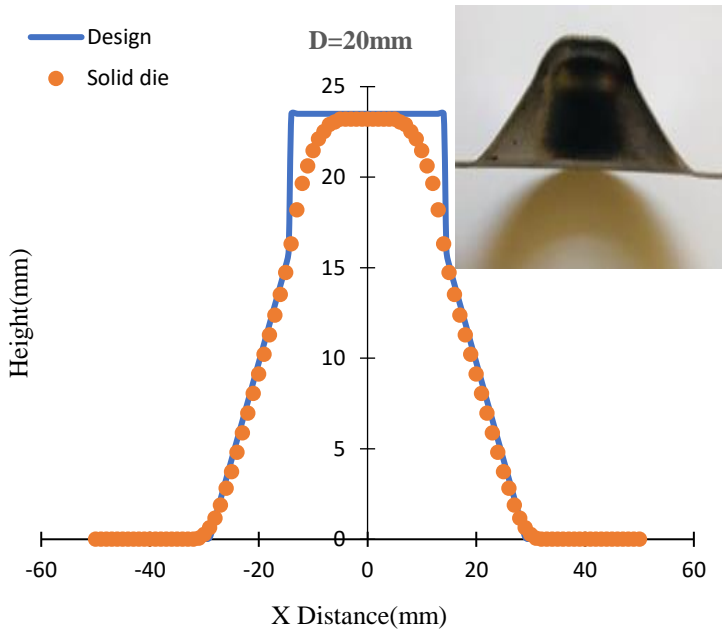

Fig.11 Comparison diagram of the effect of tool diameter of 20 mm on the dimensional accuracy of the production sample with the designed sample

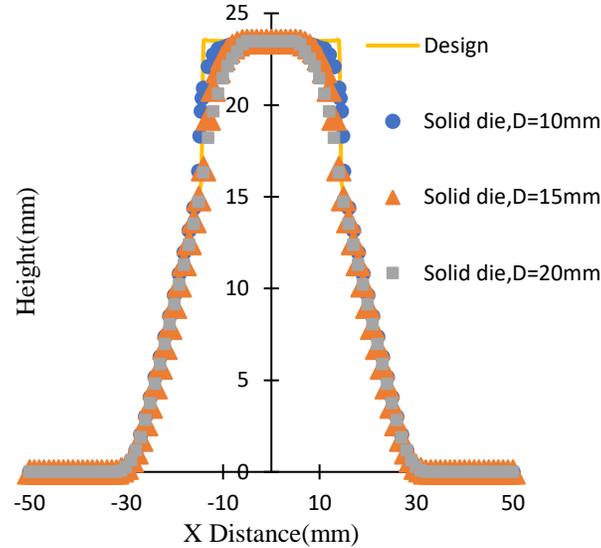

Fig.12 The diagram of the effect of the tool diameter on the dimensional accuracy

## 4.3 Analysis of the Effect of Vertical Stage on Fixed Tool Diameter on Thickness Distribution

This section compares the thickness distribution of the sheet in Experiments 2, 3, and 4, where the tool diameter is fixed at 15 mm and the vertical stages are 0.5 mm, 1 mm, and 1.5 mm, respectively. A spiral toolpath strategy is employed for forming in these experiments.

As observed in the figure13, the vertical stage within the examined range of this study has a minimal impact on the overall thickness distribution of the formed specimen. However, a closer examination of the minimum thickness region reveals that increasing the vertical stage initially leads to an increase and then a decrease in the minimum thickness.

This behavior can be attributed to the reduced number of contacts between the material and the tool in larger vertical stages. This reduction results in more unformed areas (manifested as surface roughness), consequently amplifying the bending effect. As a result, deformation is more intense in smaller vertical stages compared to larger ones, leading to more thinning and thickening, albeit to a lesser extent. Furthermore, it can be concluded that increasing the vertical stage also enhances surface roughness.

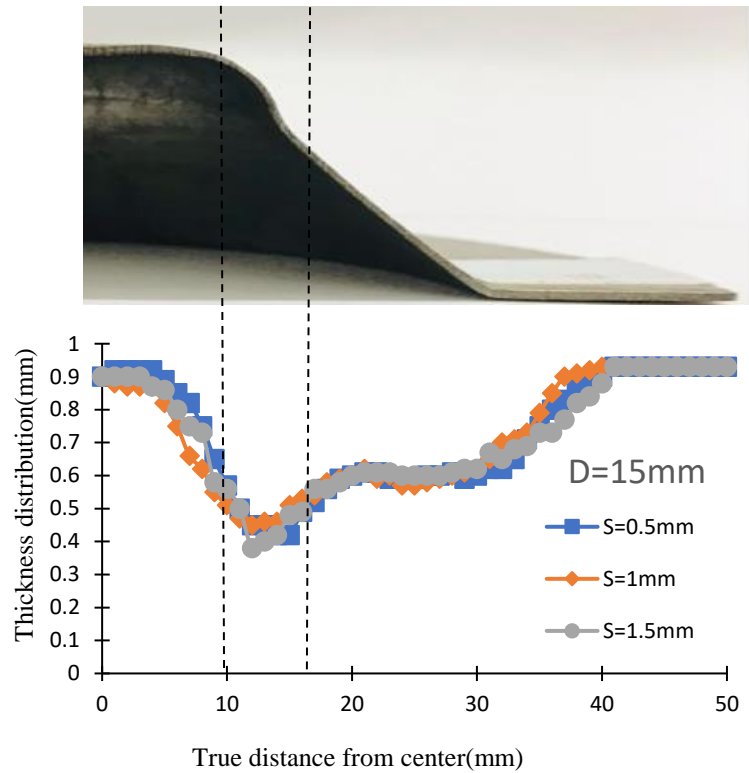

Fig.13 The diagram of the effect of the Stage down on the thickness distribution

In addition, it is noteworthy that increasing the vertical stage results in an increase in the force applied to the sheet. The reason for this is that as the vertical stage increases, the tool applies greater localized strain to the sheet. As illustrated in Figure 13 The highest minimum thickness occurs at the largest vertical stage (1.50 mm).

## 4.4 Impact of Vertical Stage Size on Dimensional Accuracy with Constant Tool Diameter

In this stage of the research, the dimensional accuracy of the sheet metal is compared across experiments 2, 3, and 4. The tool diameter is kept constant at 15 mm, while the vertical stage sizes are respectively 0.5, 1, and 1.5 mm. The experiments employ a helical movement strategy for forming.

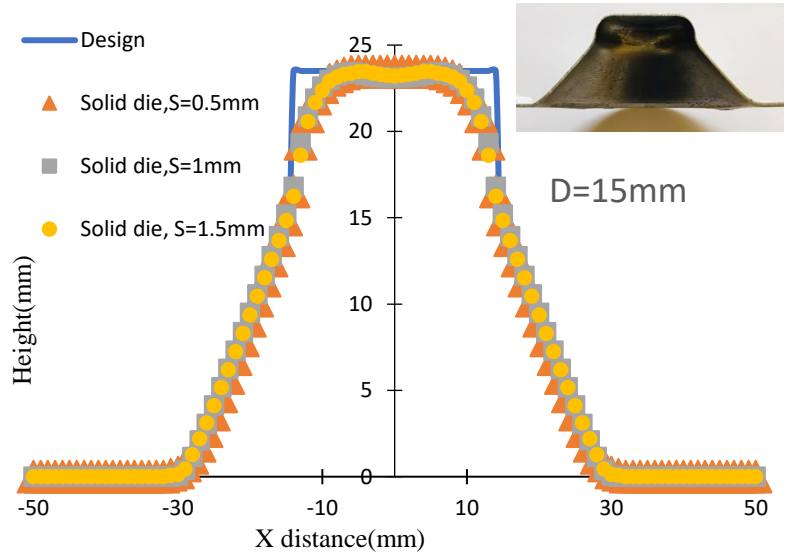

Fig.14 The diagram of the effect of the Stage down on the dimensional accuracy

As can be seen in Figure 14, variations in the vertical stage size with a constant diameter do not have a significant impact on the dimensional accuracy of the produced part.

## 4.5 Impact of Forming strategy on Dimensional Accuracy and Thickness Distribution with Constant Tool Diameter and Vertical Stage

In this stage of the research, the dimensional accuracy and thickness distribution of the produced sample are examined in experiments 1 and 7. Both experiments have a vertical stage of 0.5 mm and a tool diameter of 10 mm, with the variable parameter being the forming strategy. The forming strategies are single-stage and two-stage, respectively.

In the two-stage forming strategy (experiment 7), as shown in Figure 15, the first stage involves forming the sheet along the 47-degree angle of the die's inclined wall to the end of the die. Consequently, it is observed that the sheet is formed up to the intersection with the cone's axis of rotation (i.e., a height of 31.21 mm) without tearing. In the second stage, following the strategy and dimensions used in the other experiments, the sheet is formed to a height of 23.5 mm. Figure 16 shows the sample produced in the first stage of the two-stage forming strategy. As seen in Figure 17, the minimum thickness in the two-stage forming strategy is 0.39 mm, whereas in the single-stage forming strategy, it is 0.24 mm.

As a result, it can be observed that by employing the new forming strategy and increasing the number of stages from one to two, the minimum thickness increased. This improvement in thickness distribution reduces thinning and ultimately decreases the likelihood of tearing in the sample. It is noteworthy that the greater reduction in thickness at the center of the sample in the two-stage strategy compared to the single-stage strategy is due to the increased forming depth from 23.5 mm to 31.21 mm.

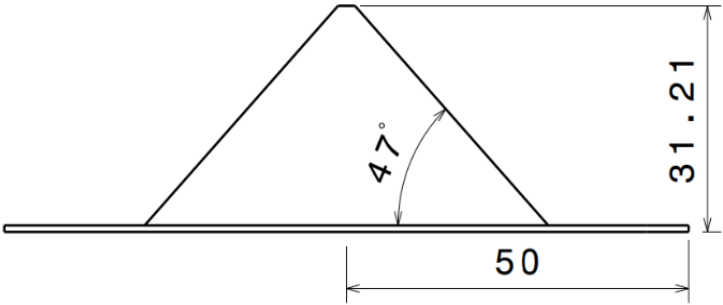

Fig.15 A view of the dimensions of the sample in the formation of the first stage of the two-stage strategy

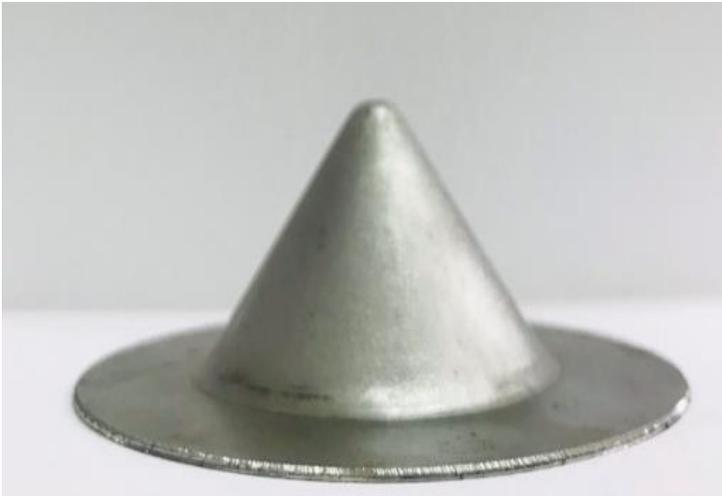

Fig.16 A view of the sample produced in the first stage of the two-stage strategy

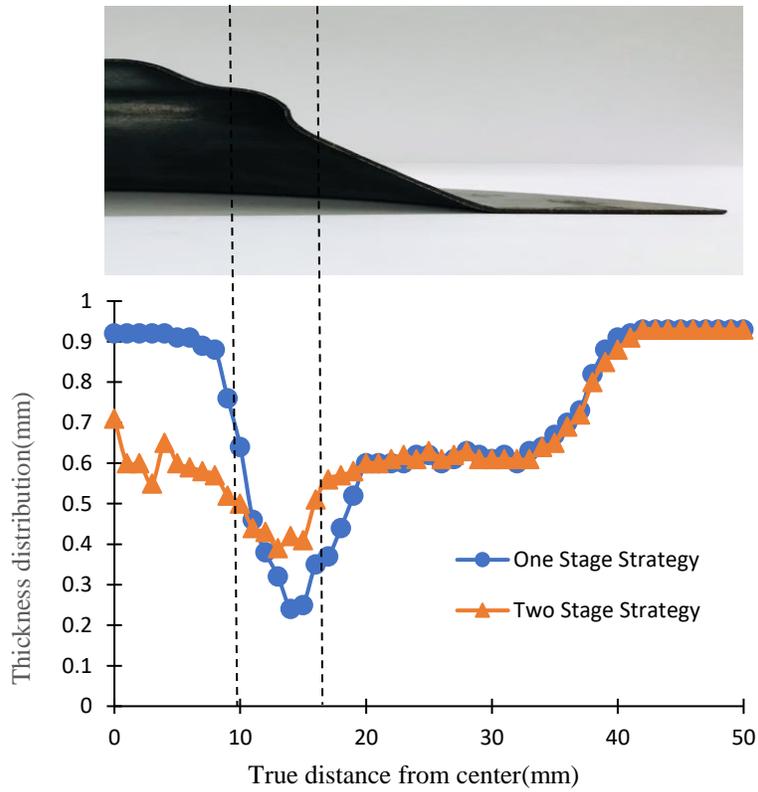

Fig.17 The diagram to investigate of the the effect of shape change strategy on thickness distribution

As shown in Figure 18, changing the forming strategy does not significantly impact the dimensional accuracy of the produced sample.

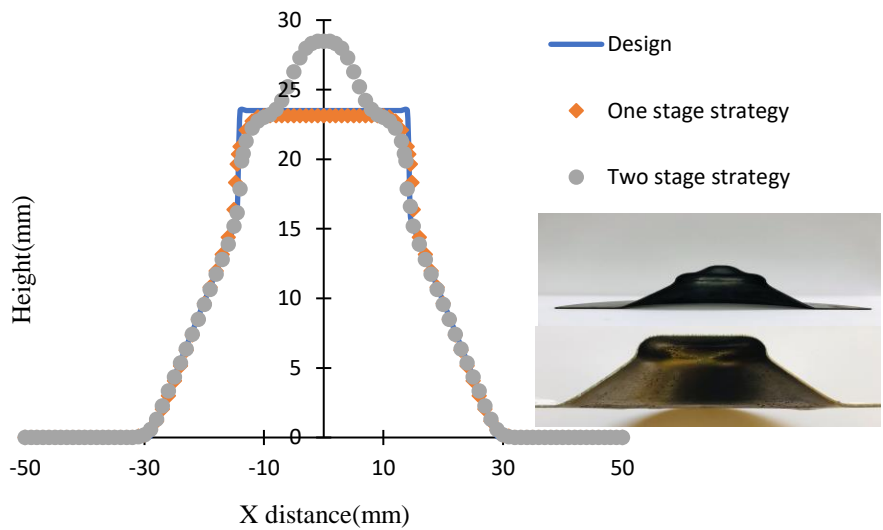

Fig.18 The diagram to investigate of the effect of shape change strategy on dimensional accuracy

## 4.6 Impact of Lubricant on Thickness Distribution and Dimensional Accuracy with Constant Tool Diameter and Vertical Stage

In this stage of the research, the dimensional accuracy and thickness distribution of the produced sample are examined in experiments 1 and 6. Both experiments have a vertical stage of 0.5 mm and a tool diameter of 10 mm, with the variable parameter being the use of lubricant. Using lubricant reduces friction between the contact surface of the tool and the sheet, which in turn decreases thinning and ultimately reduces the likelihood of tearing.

According to Figures 19 and 20, in this study, the lubricant does not significantly impact the thickness distribution and dimensional accuracy of the produced sample. It is only notable that the use of lubricant in this research primarily resulted in a reduction of surface roughness.

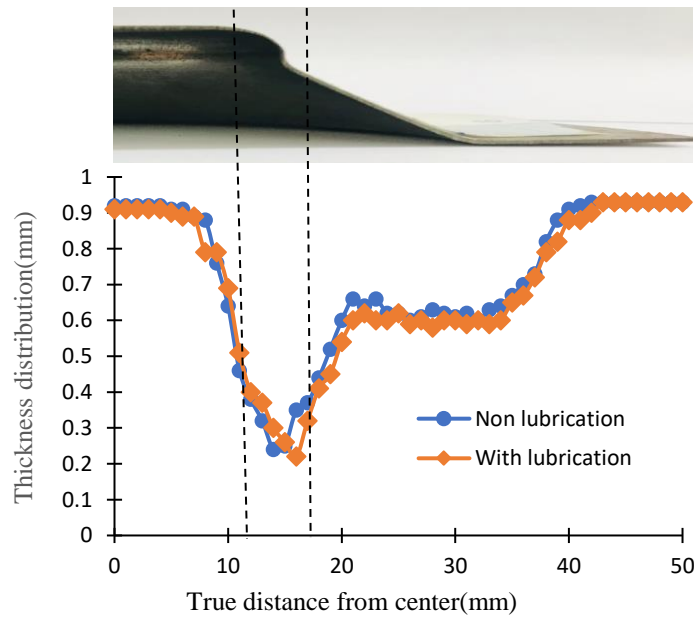

Fig.19 The diagram of the effect of the lubrication on the thickness distribution

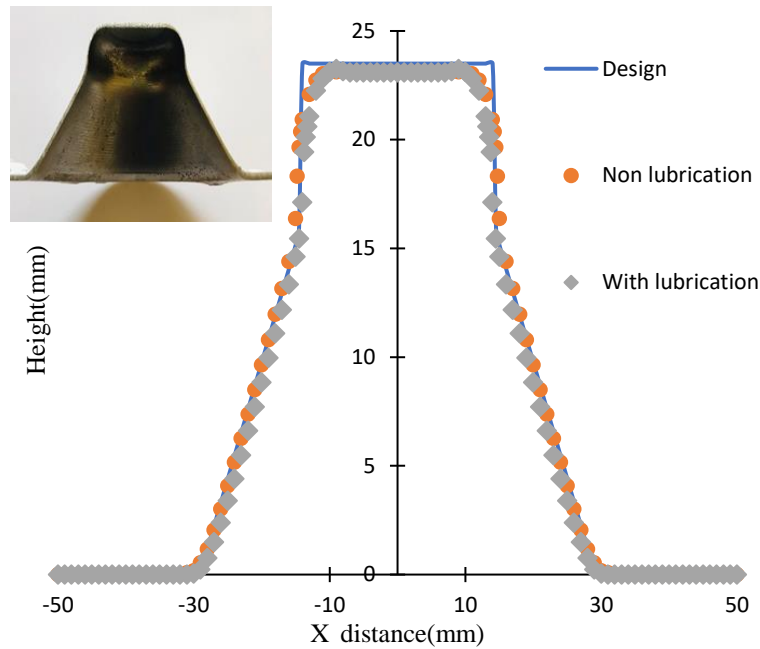

Fig.20 The diagram of the effect of the lubrication on the dimensional accuracy

## 4.7 Investigation of the Effect of Forming Strategy on Sheet Formability

In this stage of the research, the formability of the produced sample is examined in experiments 1 and 7. Both experiments have a vertical stage of 0.5 mm and a tool diameter of 10 mm, with the variable parameter being the forming strategy. The forming strategies are single-stage and two-stage, respectively. The stages for the two-stage forming strategy are performed as described at the before.

As shown in Figures 21 and 22, the maximum equivalent plastic strain in the two-stage forming strategy is 2.52, while the maximum equivalent plastic strain in the single-stage forming strategy is 2.93. Therefore, it can be observed that by employing the new forming strategy and increasing the number of stages from one to two, the maximum equivalent plastic strain decreases, which results in reduced thinning and ultimately lowers the likelihood of tearing in the sample. Hence, it can be concluded that based on the conducted investigations, increasing the number of forming stages in this scenario reduces the risk of thinning and tearing in the produced sample.

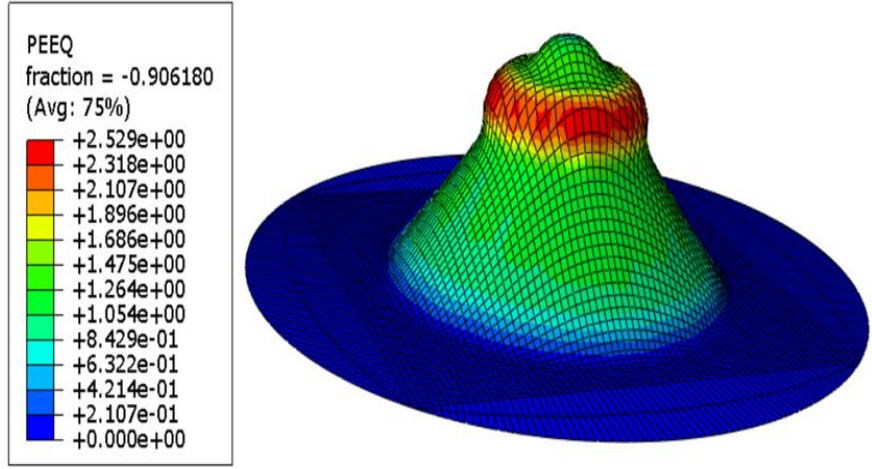

Figure 21: View of How Plastic Strain Changes in the Two-Stage forming Strategy

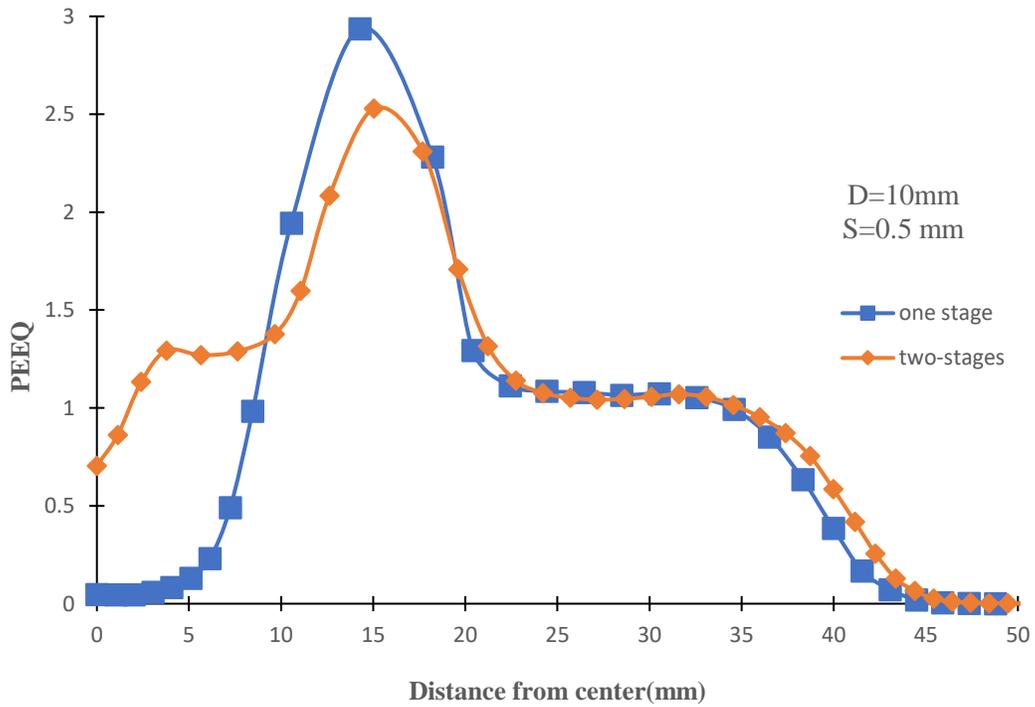

Figure 22: Diagram of the Effect of Forming strategy on Equivalent Plastic Strain

## 4.8 Investigation of the Effect of Lubricant on Formability

In this stage of the research, the formability of the produced sample is examined in experiments 1 and 6. Both experiments have a vertical stage of 0.5 mm and a tool diameter of 10 mm, with the variable parameter being the lubricant, and both are conducted in a single stage. The use of lubricant reduces friction between the tool and sheet contact surfaces, which helps decrease thinning and ultimately reduces the likelihood of tearing. However, according to Figure 23, the lubricant does not have a significant impact on the maximum equivalent plastic strain of the produced sample in this research.

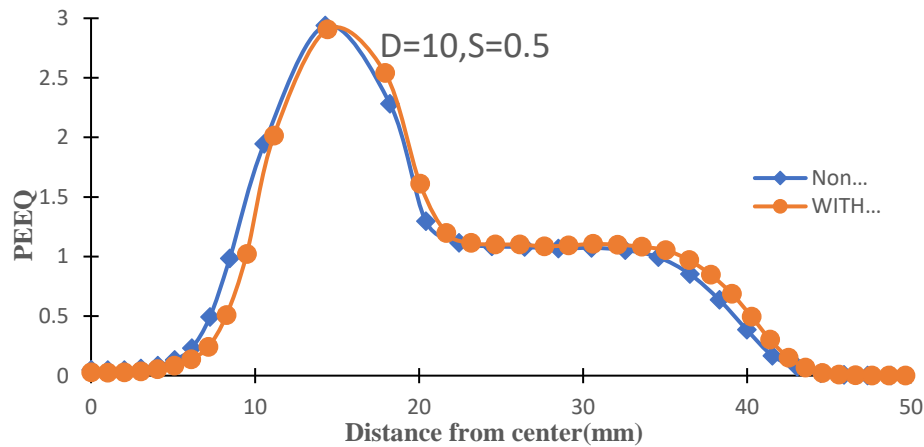

Figure 23: Diagram of the Effect of Lubricant on Equivalent Plastic Strain

## 5. Conclusion

This study investigated the incremental sheet forming (ISF) of St12 steel sheets using a hemispherical tool under various process conditions, including changes in tool diameter, lubricant type, vertical step size, and forming strategy. A negative die setup was utilized to enhance form accuracy. Both experimental and numerical analyses were conducted to evaluate the effects of these parameters on thickness distribution and dimensional accuracy. The key findings are summarized as follows:

- Increasing the tool diameter from 10 mm to 20 mm at a constant vertical step size expanded the contact area between the tool and the sheet, which improved force distribution. This resulted in an increase in minimum wall thickness from 0.24 mm to 0.50 mm, thereby reducing thinning and minimizing the risk of tearing.

- Reducing the tool diameter from 20 mm to 10 mm led to better definition of sharp corners, improving the dimensional accuracy of the formed parts.
- Increasing the vertical stage size from 0.5 to 1.5 mm with a 15 mm tool diameter initially increases and then decreases the minimum thickness. This results in increased surface roughness and bending in the produced sample.
- With a vertical stage size of 0.5 mm and a tool diameter of 10 mm, employing a new forming strategy and increasing the number of stages from one to two increases the minimum thickness from 0.24 mm to 0.39 mm. This improves the thickness distribution and reduces the likelihood of tearing. It is also noteworthy that changes in the forming strategy have no significant impact on the dimensional accuracy of the produced sample.
- According to the analyses conducted in this study, the lubricant has no significant effect on the thickness distribution and dimensional accuracy of the produced sample. The use of lubricant only resulted in a reduction of surface roughness.

- The two-stage forming strategy also resulted in a reduction in maximum equivalent plastic strain, contributing to lower thinning and improved structural integrity.

Overall, the results of this study offer practical insights for optimizing negative TPIF parameters to achieve better quality, higher formability, and enhanced reliability in incremental sheet forming applications.